\documentclass[aps,pre,nofootinbib,reprint]{revtex4-2}
\pdfoutput=1
\usepackage{amsmath,amssymb,bm,bbm,booktabs,multirow,graphicx,xspace}
\usepackage[colorlinks=true, pdfstartview=FitH, linkcolor=blue, citecolor=blue, urlcolor=blue]{hyperref}
\usepackage[capitalize]{cleveref}
\Crefname{section}{Sec.}{Secs.}
\Crefname{figure}{Fig.}{Figs.}

\graphicspath{{./plots/}}

\renewcommand\phi\varphi
\newcommand\ev[1]{\left\langle#1\right\rangle}

\newcommand\pn{\tilde P^{(N)}}
\newcommand\AId{AI$^\dagger$\xspace}
\newcommand\AIId{AII$^\dagger$\xspace}
\newcommand\z{v}
\newcommand\w{w}

\begin{document}

\title{Approximation formula for complex spacing ratios in the Ginibre ensemble}

\author{Ioachim G. Dusa and Tilo Wettig}
\affiliation{Department of Physics, University of Regensburg, 93040 Regensburg, Germany}

\begin{abstract}
  Recently, S\'a, Ribeiro, and Prosen introduced complex spacing ratios to analyze eigenvalue correlations in non-Hermitian systems. At present there are no analytical results for the probability distribution of these ratios in the limit of large system size. We derive an approximation formula for the Ginibre universality class of random matrix theory which converges exponentially fast to the limit of infinite matrix size. We also give results for moments of the distribution in this limit.
\end{abstract}

\maketitle

\section{Introduction}

The theoretical description of a given physical system typically involves a certain operator. For example, in quantum systems this could be the Hamilton operator or the Dirac operator. The eigenvalues of this operator, collectively called its spectrum, carry important information on the properties of the system. Naturally, the eigenvalues depend on the details of the system. However, in sufficiently complex systems, spectral correlations on the scale of the mean level spacing turn out to be universal; i.e., they do not depend on the detailed dynamics of the system but only on global (antiunitary) symmetries \cite{10.2307/1969342, 10.1063/1.1703863}. It is well known that random matrix theory (RMT) can be used to compute these universal correlations (see, e.g., \cite{Mehta:2004, Guhr:1997ve, Verbaarschot:2000dy, Oxford:2015}).

To compare results for a given physical system to RMT, the universal and nonuniversal features of the spectrum need to be separated. This is usually done by a so-called unfolding procedure, which attempts to fit the global spectral density. There is some ambiguity in the choice of this procedure. Fortunately, unfolding can be avoided by considering the ratio of consecutive eigenvalue spacings \cite{PhysRevB.75.155111, PhysRevLett.110.084101}, for which the local spectral density drops out. Recently, this idea has been generalized to the non-Hermitian case with complex eigenvalues. Specifically, S\'a, Ribeiro, and Prosen \cite{PhysRevX.10.021019} proposed to study the probability distribution $P(z)$ of the quantity
\begin{align}
  \label{eq:csr}
  z_k=\frac{\lambda_k^\text{NN}-\lambda_k}{\lambda_k^\text{NNN}-\lambda_k}\,,
\end{align}
where $\lambda_k^\text{NN}$ and $\lambda_k^\text{NNN}$ are the nearest and next-to-nearest neighbors of $\lambda_k$ in the complex plane. This avoids unfolding in the complex plane, which is more difficult than for real spectra \cite{Markum:1999yr, PhysRevLett.123.254101}. 

The classification of RMT involves 10 symmetry classes in the Hermitian case \cite{Zirnbauer:1996zz, Altland:1997zz} and 38 symmetry classes in the non-Hermitian case \cite{Kawabata:2018gjv, PhysRevB.99.235112}. However, for the bulk spectral correlations there are only three \emph{universality classes} in both cases: GOE, GUE, and GSE in the Hermitian case \cite{10.1063/1.1703863}, and Ginibre, \AId, and \AIId in the non-Hermitian case \cite{Ginibre:1965zz, Jaiswal_2019, PhysRevResearch.2.023286}. In the Hermitian case, Wigner-like surmises for the spacing ratios have been computed in \cite{PhysRevLett.110.084101} for all three universality classes. They are obtained from $3\times3$ matrices and give very good approximations to the corresponding $N\to\infty$ limits, where $N$ is the matrix size. In the non-Hermitian case, a Wigner-like surmise obtained from $3\times3$ matrices has been computed for the Ginibre unitary ensemble in \cite{PhysRevX.10.021019}, but it was found to differ significantly from the $N\to\infty$ limit. This is largely due to boundary effects, which led the authors of \cite{PhysRevX.10.021019} to consider random-matrix ensembles with smaller boundary effects but the same large-$N$ distribution, such as the toric unitary ensemble. Indeed, the numerical results obtained in this ensemble for small $N$ show much better agreement with the large-$N$ limit, but the agreement is not as good as for the Hermitian Wigner-like surmises, and simple analytical results are not available either.

Our aim in this paper is to derive a simple finite-$N$ approximation to $P(z)$ that converges rapidly to the $N\to\infty$ limit. We will show that our approximation converges to this limit  exponentially fast in $N$, while the exact finite-$N$ result only converges like $1/\sqrt N$. We restrict ourselves to the Ginibre universality class, which is relevant for most non-Hermitian systems. (Note that the short-range bulk spectral correlations are identical in the Ginibre orthogonal, unitary, and symplectic ensembles \cite{PhysRevLett.123.254101}.) The universality classes \AId and \AIId are nonstandard but still have interesting applications \cite{Jaiswal_2019, PhysRevResearch.2.023286, Kanazawa:2021zmv, garciagarcia2021symmetry}. To the best of our knowledge, no attempt to compute $P(z)$ for these two universality classes has been made in the literature yet.

This paper is organized as follows. In \cref{sec:derivation} we derive our approximation formula for $P(z)$. In \cref{sec:Pz} we compute the corresponding marginal distributions $P(r)$ and $P(\theta)$, where $z=re^{i\theta}$. In \cref{sec:numerics} we present numerical results, both for our approximation formula and for RMT at finite $N$, to demonstrate the convergence behavior. We also give reference results for several moments of $P(z)$ in the $N\to\infty$ limit. We conclude in \cref{sec:summary}. Our computer algebra code is publicly available \cite{csratio}.

\section{Derivation of the approximation}
\label{sec:derivation}

The groundwork for our approximation formula has already been laid in \cite{PhysRevX.10.021019}.  We start directly from Eq.~(C6) of that reference, which gives the exact distribution $P^{(N)}(z)$ for finite $N$. Since we are interested in the large-$N$ limit we neglect terms of order $1/N$ in the exponential terms and obtain the approximation
\begin{widetext}
  \begin{align}
  \pn(x,y)
  &\propto\Theta(1-(x^2+y^2))(x^2+y^2)(1+x^2+y^2-2x)\int ds\,dt\,(s^2+t^2)^4\exp\left\{-(s^2+t^2)(1+x^2+y^2)\right\}\notag\\
  &\quad\times I_{N-3}(x,y,s,t)
    \label{eq:rhoN}
\end{align}
with $z=x+iy$ and
\begin{align}
  I_N(x,y,s,t)
  &=\int \prod_{j=1}^Nda_j\,db_j\,\Theta((a_j^2+b_j^2)-(s^2+t^2))(a_j^2+b_j^2)\left[(a_j-s)^2+(b_j-t)^2\right]\notag\\
  &\quad\times\left[(a_j-sx+ty)^2+(b_j-tx-sy)^2\right]\exp\left\{-(a_j^2+b_j^2)\right\}\prod_{j<k}^N\left[(a_j-a_k)^2+(b_j-b_k)^2\right].
    \label{eq:IN}
\end{align}
\end{widetext}
All integrations in these formulas go from $-\infty$ to $+\infty$. We emphasize that $\pn(x,y)$ as given above is not the exact result for a given $N$ but an approximation that improves as $N$ becomes large. We see in Sec.~\ref{sec:numerics} that this approximation converges to the $N\to\infty$ limit $P(z)$ exponentially fast in $N$, while the exact finite-$N$ result $P^{(N)}(z)$ converges to $P(z)$ only like $1/\sqrt N$.

The integrand of $I_N$ factorizes except for the double product at the end. To factorize this term we introduce complex variables $\z_j=a_j+ib_j$ and employ standard methods (see, e.g., Sec.~8.8.1 of \cite{Haake2010}) to rewrite it as
\begin{align}
  \prod_{j<k}^N&\left[(a_j-a_k)^2+(b_j-b_k)^2\right]
  =\prod_{j<k}^N\left|\z_j-\z_k\right|^2\notag\\
  &=\det\biggl(\sum_{j=1}^N\z_j^{i-1}\z_j^{*k-1}\biggr)_{i,k=1,\ldots,N}\notag\\
  &=\int \prod_{j=1}^N d\eta_j\,d\eta_j^*\prod_{j=1}^N\biggl(\sum_{i,k=1}^N\eta_i^*\eta_k\z_j^{i-1}\z_j^{*k-1}\biggr),
\end{align}
where in the last term the integral is over anticommuting variables. Furthermore, we write another part of the integrand of $I_N$ in the form
\begin{multline}
  \lefteqn{(a_j^2+b_j^2)\left[(a_j-s)^2+(b_j-t)^2\right]}\\
  \times\left[(a_j-sx+ty)^2+(b_j-tx-sy)^2\right]\\
  =\sum_{m,n=1}^3M_{mn}(x,y,s,t)\z_j^m\z_j^{*n}
\end{multline}
with coefficient functions $M_{mn}(x,y,s,t)$ that satisfy $M_{mn}=M_{nm}^*$ and can easily be determined by inspection [see Eqs.~\eqref{eq:Mmn} below]. This gives
\begin{align}
  I_N&(x,y,s,t)=\int \prod_{j=1}^N d\eta_j\,d\eta_j^*\notag\\
  \times&\biggl[\sum_{i,k=1}^N\eta_i^*\eta_k\sum_{m,n=1}^3M_{mn}(x,y,s,t)J_{m+i,n+k}(s^2+t^2)\biggr]^N
\end{align}
with
\begin{align}
  J_{p,q}(x)=\int_{|\z|^2>x}d^2\z\,\z^{p-1}\z^{*q-1}e^{-|\z|^2}=\pi\Gamma(p,x)\delta_{pq}\,,
\end{align}
where in the last step we observed that the integral over the complex variable $\z$ is nonzero only for $p=q$ and used the definition of the incomplete $\Gamma$ function,
\begin{align}
  \Gamma(p,x)=\int_x^\infty du\,u^{p-1}e^{-u}\,.
\end{align}
We now introduce the complex variable $\w=s+it$ and define the $N$-dimensional pentadiagonal matrix
\begin{align}
  A_{ik}=\sum_{m,n=1}^3M_{mn}\Gamma(m+i,|\w|^2)\delta_{m+i,n+k}\,,
\end{align}
which depends on $x$, $y$, and $\w$ and is Hermitian because of $M_{mn}=M_{nm}^*$. This yields
\begin{align}
  I_N=\pi^N\int \prod_{j=1}^N d\eta_j\,d\eta_j^*\left(\eta^\dagger A\eta\right)^N=\pi^NN!\det A\,.
\end{align}

At this point \cref{eq:rhoN} has been reduced to a two-di\-men\-sio\-nal integral over $\w$.
In terms of $\w$ the $M_{mn}$ are given by
\begin{align}
  M_{11}&=|\w|^4(x^2+y^2)\,,\;\: M_{12}\hspace*{-1.1em} &&=-\w|\w|^2(x^2+y^2+x+iy)\,,\notag\\
  M_{13}&=\w^2(x+iy)\,, && M_{22}=|\w|^2((1+x)^2+y^2)\,,\notag\\
  M_{23}& =-\w(1+x+iy)\,, && M_{33}=1\,.
  \label{eq:Mmn}
\end{align}
Writing $\w=\sqrt{R}e^{i\phi}$ and using the explicit expressions for the $M_{mn}$ it is straightforward to show by induction that $\det A$ depends only on $R$ so that the integral over $\phi$ becomes trivial. We thus obtain
\begin{align} 
  \pn(x,y)&=\Theta(1-(x^2+y^2))(x^2+y^2)((1-x)^2+y^2)\notag\\
  &\quad \times K_{N-3}(x,y)
\end{align}
with
\begin{align}
  \label{eq:KN}
  K_N(x,y)=c_N\int_0^\infty dR\,R^4e^{-R(1+x^2+y^2)}\det A\,,
\end{align}
where $\det A$ depends on $x$, $y$, and $R$, and $c_N$ is a normalization constant which ensures that $\pn(x,y)$ integrates to $1$. The computation of $\pn(z)$ has now been reduced to the evaluation of the one-dimensional integral in \cref{eq:KN}. One can show from the structure of $A$ that the integral converges. The integral is suitable for numerical evaluation. However, we can perform further analytical manipulations, which lead to manageable approximation formulas in terms of power series.

Using the definition of $A$ and the fact that the first argument of the incomplete $\Gamma$ function is an integer one can show that $\det A=P(R)e^{-NR}$, where $P(R)=\sum_{k=0}^{N(N+5)/2}c_kR^k$ is a polynomial in $R$ with coefficients $c_k$ that are polynomials in $x$ and $y$. The individual terms in $P(R)$ can be integrated analytically to give
\begin{align}
  K_N(x,y)=c_N\sum_{k=0}^{N(N+5)/2}\frac{c_k(x,y)(k+4)!}{(N+1+x^2+y^2)^{k+5}}\,.
\end{align}
In the following we write $x+iy=z=re^{i\theta}$ with $r\in[0,1]$ and $\theta\in[0,2\pi]$. From the structure of the $M_{mn}$ and the fact that $\det A$ is real one can show that only even powers of $y$ appear in $c_k(x,y)$. Thus, the latter can be written as polynomials in $r$ and $\cos\theta$. Pulling out the leading factor in the denominator, we obtain
\begin{align}
  \label{eq:cmn}
  K_N(r,\theta)&=\frac1{[1+r^2/(N+1)]^{\frac12N(N+5)+5}}\notag\\
  &\quad\times\sum_{m=0}^{N(N+5)}\sum_{n=0}^Nc_{mn}r^m\cos^n\theta\,,
\end{align}
where the coefficients $c_{mn}$ are zero if $m+n$ is odd, which can be seen by writing $r^m\cos^n\theta=(x^2+y^2)^{(m-n)/2}x^n$. It is straightforward to determine the $c_{mn}$ by computer algebra, and we provide a \textsc{sympy} program for this purpose~\cite{csratio}. Our approximate result thus reads
\begin{align}
  \label{eq:Pz}
  \pn(r,\theta)=\Theta(1-r)r^3(r^2-2r\cos\theta+1)K_{N-3}(r,\theta)\,,
\end{align}
where the additional factor of $r$ on the right-hand side is the Jacobian of the transformation from $(x,y)$ to $(r,\theta)$.

\section{Marginal distributions}
\label{sec:Pz}

For practical applications it is useful to consider the marginal distributions $P(r)$ and $P(\theta)$, which are obtained by integrating $P(r,\theta)$ over $\theta$ and $r$, respectively. For the integration over $\theta$ we use
\begin{align}
  \int_0^{2\pi}d\theta\,\cos^n\theta=
  \begin{cases}
    2\pi\frac{(n-1)!!}{n!!}\,, & n\text{ even}\,,\\
    0\,, & n\text{ odd}\,.
  \end{cases}
\end{align}
For the integration over $r$ we encounter integrals of the type
\begin{align}
  I_{n,p}(u)=\int_0^1dr\,\frac{r^n}{(u+r^2)^p}
\end{align}
with $u>0$ and nonnegative integers $n$ and $p$. These integrals can be done analytically for any $n$ and $p$, which can be accomplished using the recursion relations
\begin{subequations}
  \begin{align}
    \label{eq:rec1}
    I_{n,p+1}&=-\frac1p\frac{d}{du}I_{n,p}\,,\\
    \label{eq:rec2}
    I_{n+2,p}&=I_{n,p-1}-uI_{n,p}    
  \end{align}
\end{subequations}
with starting values
\begin{subequations}  
  \begin{align}
    \label{eq:start1}
    I_{n,0}&=\frac1{n+1}\,,\\
    \label{eq:start2}
    I_{0,1}&=\frac1{\sqrt u}\arctan\frac1{\sqrt u}\,,\qquad
    I_{1,1}=\frac12\ln\frac{u+1}u\,.
  \end{align}
\end{subequations}
We first use \cref{eq:rec1} to compute $I_{0,p}$ and $I_{1,p}$, starting from \cref{eq:start1} and going up to the desired value of $p$. Afterwards, we use \cref{eq:rec2} to compute the desired $I_{n,p}$, starting from \cref{eq:start2}.
Performing the integrations, we obtain
\begin{align}
  \label{eq:Pr}
  \pn(r)
  &=\int_0^{2\pi}d\theta\,\pn(r,\theta)\notag\\
  &=\frac{\Theta(1-r)r^3}{[1+r^2/(N-2)]^{\frac12N(N-1)+2}}
    \!\!\sum_{n=0}^{\frac12N(N-1)-2}\!\!a_nr^{2n}\,,\\
  \label{eq:Ptheta}
  \pn(\theta)&=\int_0^1dr\,\pn(r,\theta)=\sum_{n=0}^{N-2}b_n\cos^n\theta
\end{align}
with coefficients $a_n$ and $b_n$ that are again straightforward to determine by computer algebra~\cite{csratio}. From the results of Eqs.~\eqref{eq:Pz}, \eqref{eq:Pr}, and \eqref{eq:Ptheta} we can also compute various moments of these distributions, for example,
\begin{align}
  \ev{r\cos\theta}=\int_0^\infty dr\int_0^{2\pi}d\theta\,\pn(r,\theta)\,r\cos\theta\,,
\end{align}
by simply using the integrals given above.

\section{Numerical results}
\label{sec:numerics}

\subsection{\boldmath Convergence of the approximation}

\begin{figure*}
  \centering
  \begin{tabular}{r@{\hspace*{10mm}}r}
    \includegraphics[height=.195\textheight]{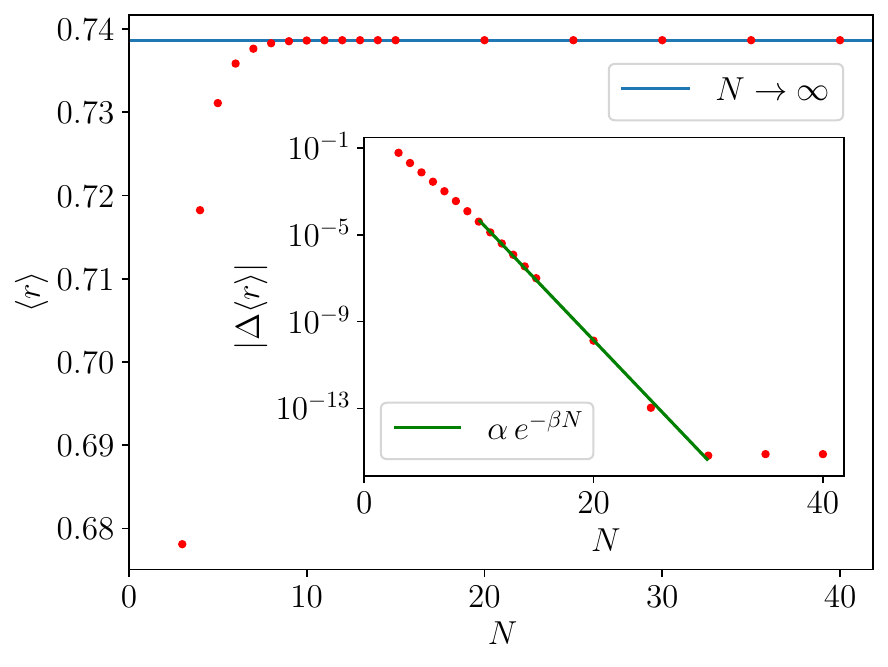} &
    \includegraphics[height=.195\textheight]{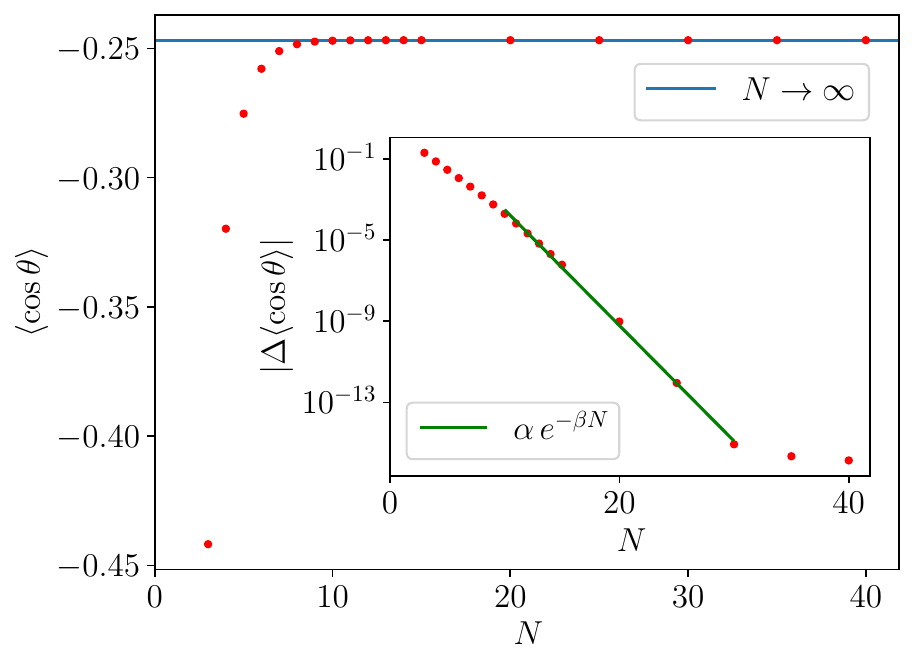} \\
    \includegraphics[height=.195\textheight]{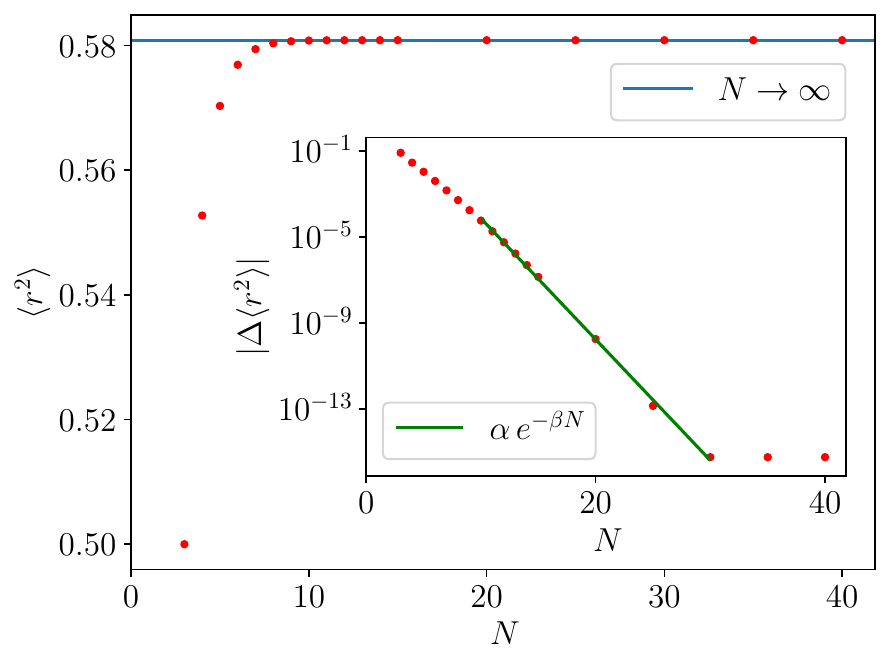} &
    \includegraphics[height=.195\textheight]{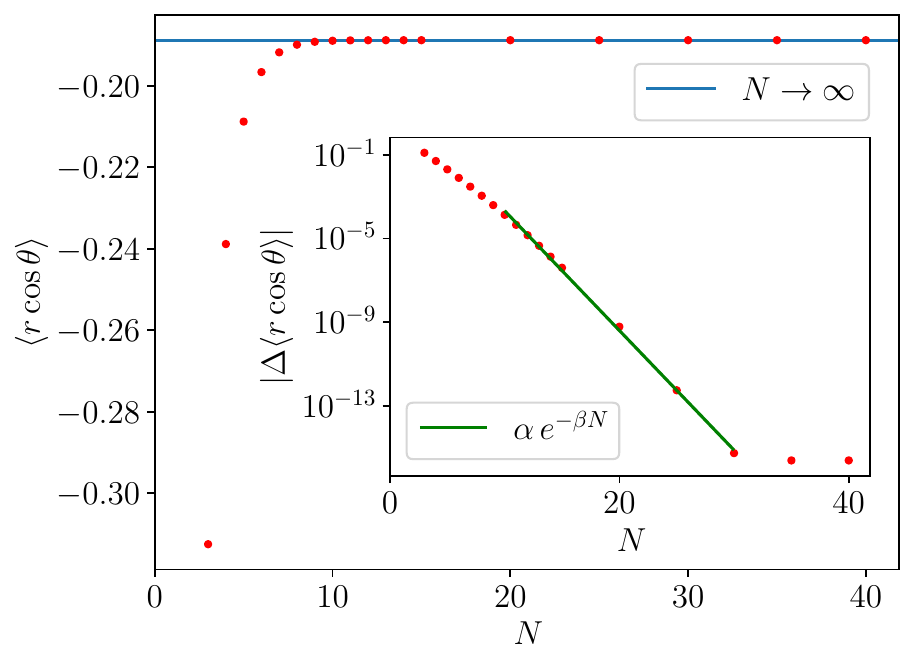}
  \end{tabular}
  \caption{Convergence of our approximation (red circles) towards the $N\to\infty$ limit (horizontal lines) for several moments of $\pn(z)$. The insets show the differences relative to the $N\to\infty$ limit. The curves in the insets are fits to an exponential in the range $N\in[10,30]$. For $N\ge30$ we observe numerical noise since our numerical precision was set to $10^{-13}$.}
  \label{fig:moments_analytical}
\end{figure*}

A very interesting question is how rapidly our approximation converges to the $N\to\infty$ limit. To answer this question we first compute various moments of $\pn(z)$ and investigate their behavior as we increase $N$. We have checked that we obtain identical results from integrating \cref{eq:KN} numerically and from using Eqs.~\eqref{eq:cmn}, \eqref{eq:Pr}, or \eqref{eq:Ptheta}, where ``identical'' refers to the relative precision of $10^{-13}$ we used in the numerical integration. In \cref{fig:moments_analytical} we show our results. We observe a very rapid convergence to the $N\to\infty$ limit,\footnote{For $N\ge30$ our results differ only by numerical noise due to finite precision. We somewhat arbitrarily took the results for $N=50$ to define the $N\to\infty$ limit. Any $N\ge20$ yields the same numbers in \cref{tab:moments}.} which appears to be exponentially fast. Already at $N=20$ the results agree with the $N\to\infty$ limit within nine significant digits. This is quite different from the behavior of large random matrices, where the convergence to the $N\to\infty$ limit goes like $1/\sqrt N$ (see \cref{fig:moments_rmt} in \cref{sec:rmt} below). The results for the $N\to\infty$ limit of the moments are given in \cref{tab:moments}.

\begin{table}[h]
  \centering
  \begin{tabular}{c@{\hspace*{2em}}r}
    \toprule
    Moment & \multicolumn{1}{c}{$N\to\infty$ limit} \\
    \midrule
    $\ev{r}$ & 0.73866010\\
    $\ev{r^2}$ & 0.58084906\\
    $\ev{\cos\theta}$ & $-$0.24683082\\
    $\ev{\cos^2\theta}$ & 0.44991388\\
    $\ev{r\cos\theta}$ & $-$0.18882912\\
    \bottomrule
  \end{tabular}
  \caption{$N\to\infty$ limit of several moments of $\pn(z)$.}
  \label{tab:moments}
\end{table}

\begin{figure*}
  \centering
  \begin{tabular}{r@{\hspace*{10mm}}r}
    \includegraphics[height=.2\textheight]{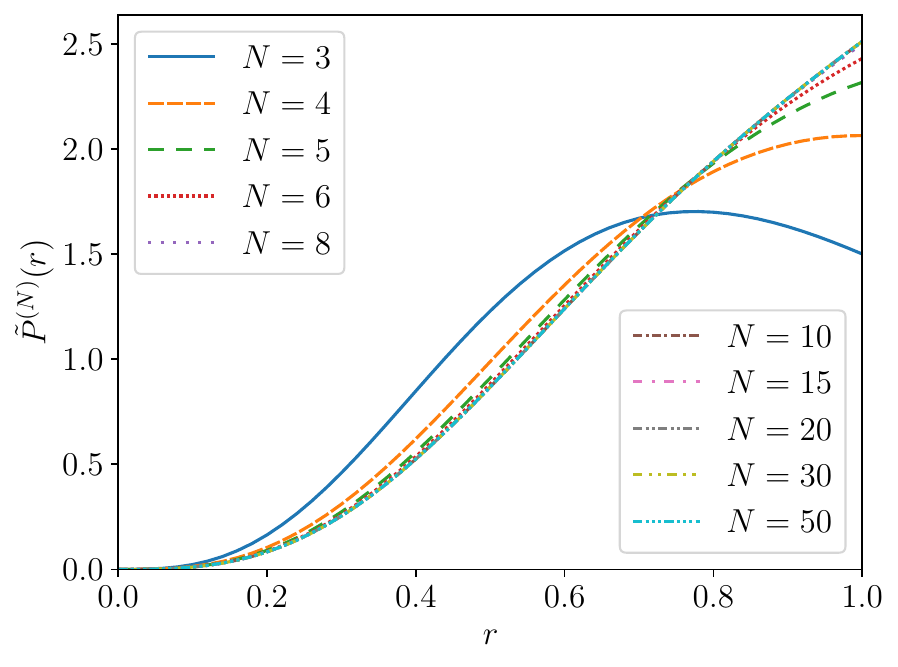} &
    \includegraphics[height=.2\textheight]{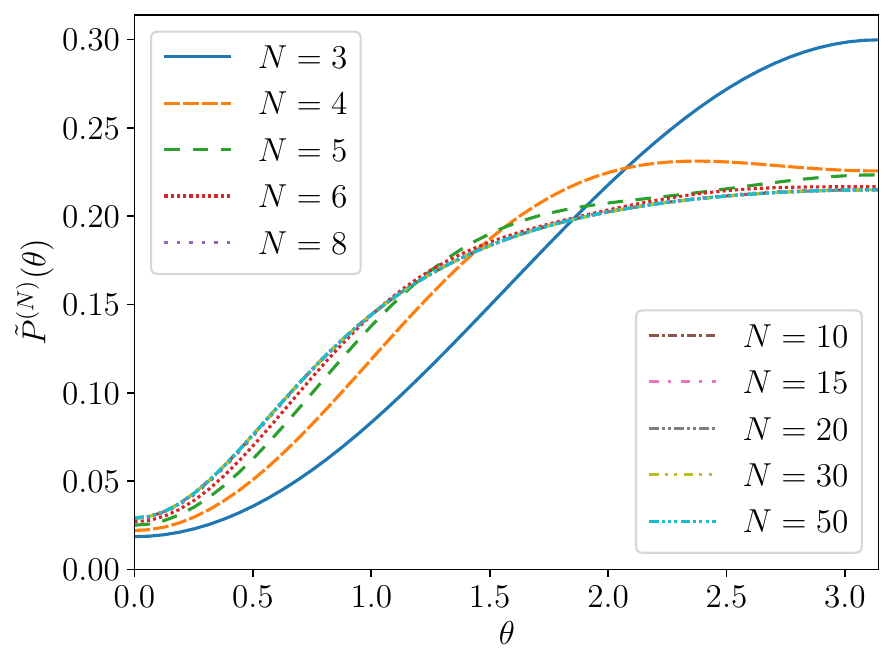} \\
    \includegraphics[height=.2\textheight]{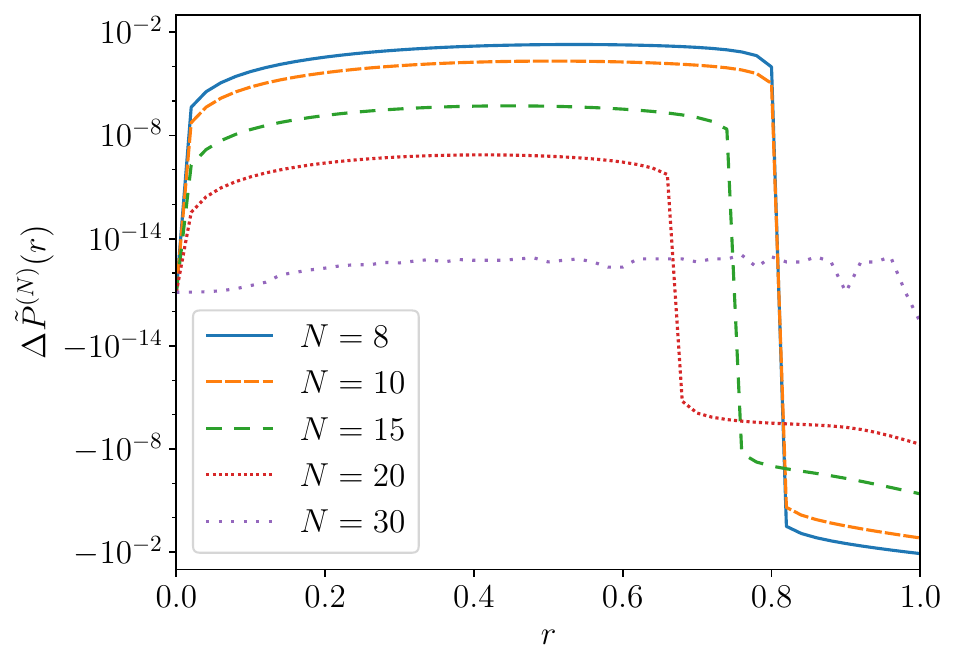} &
    \includegraphics[height=.2\textheight]{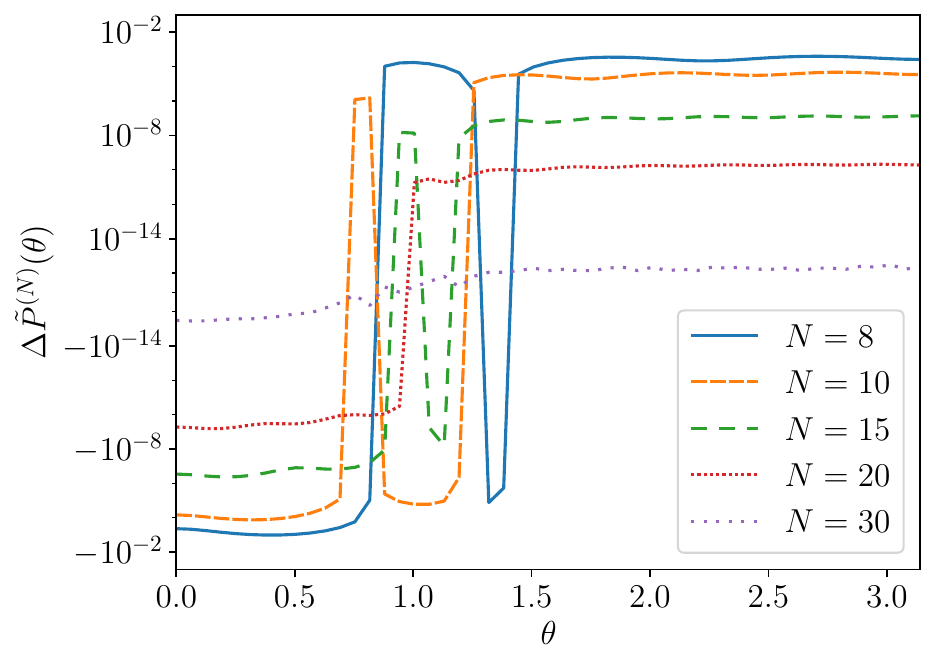} \\[-1.5mm] 
  \end{tabular}
  \caption{Top: Marginal distributions $\pn(r)$ and $\pn(\theta)$ for various values of $N$. We only show the interval $\theta\in[0,\pi]$ since $\pn(\theta)$ is invariant under $\theta\to2\pi-\theta$. Already at $N=8$ the curves are essentially indistinguishable to the naked eye from the $N\to\infty$ limit. Bottom: Differences between the marginal distributions and the corresponding $N\to\infty$ limits. Our numerical precision was set to $10^{-13}$.}
  \label{fig:P_analytical}
\end{figure*}

Given that the moments converge rapidly, we expect the distributions to do so as well. We do not show two-dimensional plots of $\pn(z)$ since it is difficult to extract information on the $N$ dependence from them. Instead, we show the marginal distributions in \cref{fig:P_analytical}. They confirm our expectations and approximate the $N\to\infty$ limit very well already for $N\sim10$. The coefficients $a_n$ and $b_n$ in \cref{eq:Pr,eq:Ptheta} are tabulated for $N=10$ in \cref{tab:anbn}.  We refer the interested reader to our computer algebra program \cite{csratio} for the case of general $N$.

\begin{table}
  \centering
  \begin{tabular}{r@{\hspace{1.8em}}r@{\hspace{1.8em}}r@{\hspace{1em}}r}
    \toprule
    $n$ & \multicolumn{1}{c}{$a_n$}
    & \multicolumn{1}{c}{$a_{n+10}$} & \multicolumn{1}{c}{$b_n$} \\
    \midrule
     0 & 1.1343071e+01 & 5.0846897e+00 &  1.8692290e$-$01 \\
     1 & 4.2370622e+01 & 1.8756273e+00 & $-$4.8290999e$-$02 \\
     2 & 8.3036308e+01 & 6.0958526e$-$01 & $-$3.5195979e$-$02 \\
     3 & 1.1290599e+02 & 1.7528443e$-$01 & $-$2.6644115e$-$02 \\
     4 & 1.1877675e+02 & 4.4809124e$-$02 & $-$1.7668758e$-$02 \\
     5 & 1.0217790e+02 & 1.0233169e$-$02 & $-$6.3894405e$-$03 \\
     6 & 7.4201384e+01 & 2.0971650e$-$03 & $-$5.9691982e$-$03 \\
     7 & 4.6398521e+01 & 3.8722534e$-$04 & $-$1.1634604e$-$02 \\
     8 & 2.5315302e+01 & 6.4636142e$-$05 & $-$6.1396515e$-$03 \\
     9 & 1.2110800e+01 & 9.7808814e$-$06 \\
    \bottomrule
  \end{tabular}
  \caption{Coefficients of the marginal distributions for $N=10$. We omitted the $a_n$ for $n\ge20$ since they decrease rapidly.}
  \label{tab:anbn}
\end{table}

\subsection{Convergence of RMT results}
\label{sec:rmt}

\begin{figure*}
  \centering
  \begin{tabular}{r@{\hspace*{10mm}}r}
    \includegraphics[height=.2\textheight]{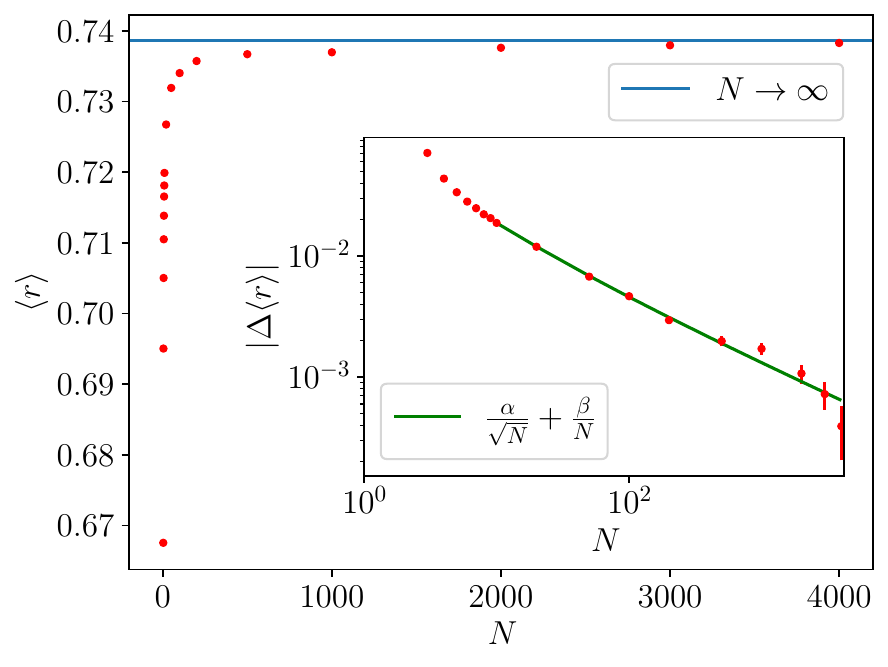} &
    \includegraphics[height=.2\textheight]{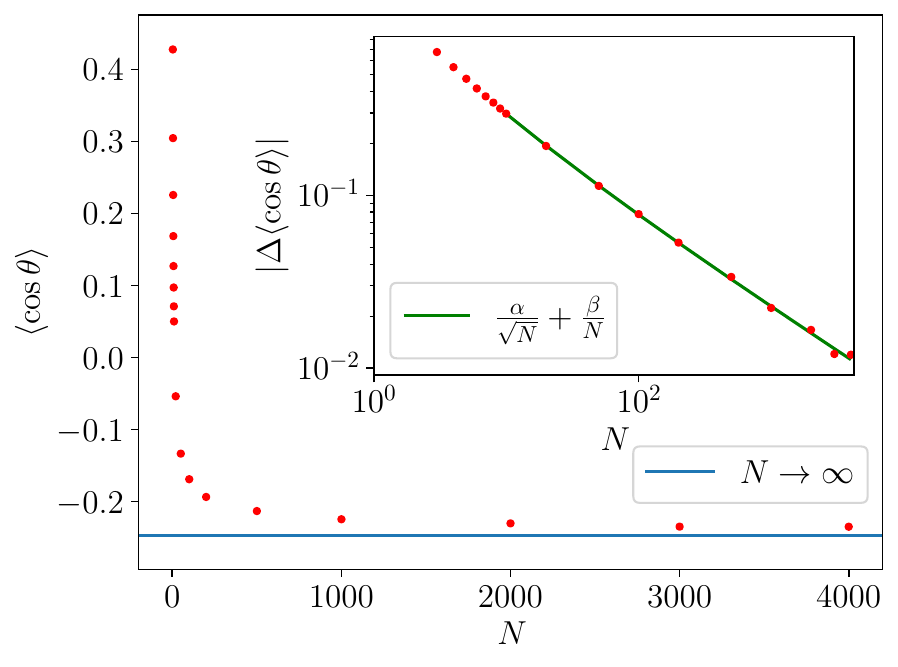} \\
    \includegraphics[height=.2\textheight]{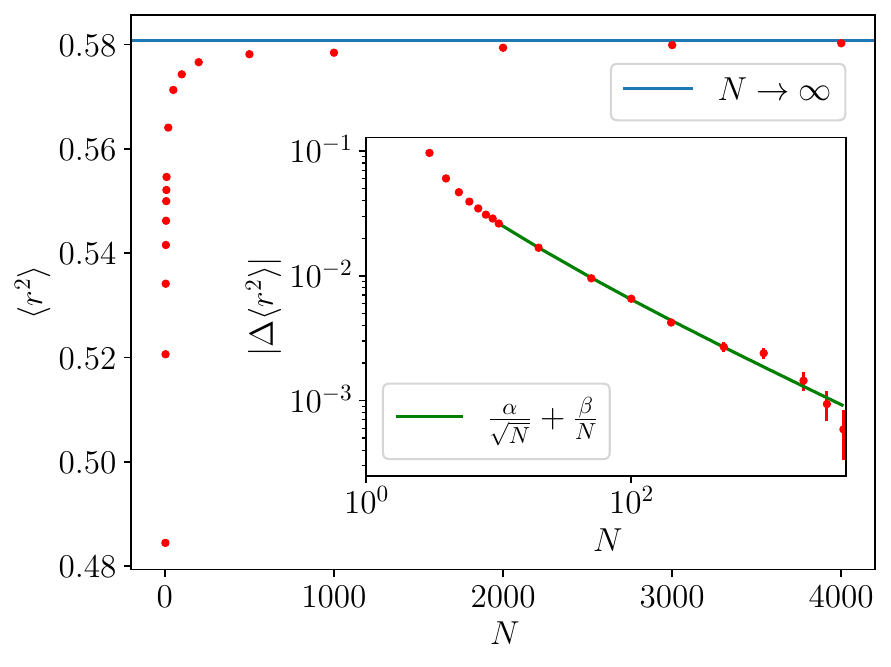} &
    \includegraphics[height=.2\textheight]{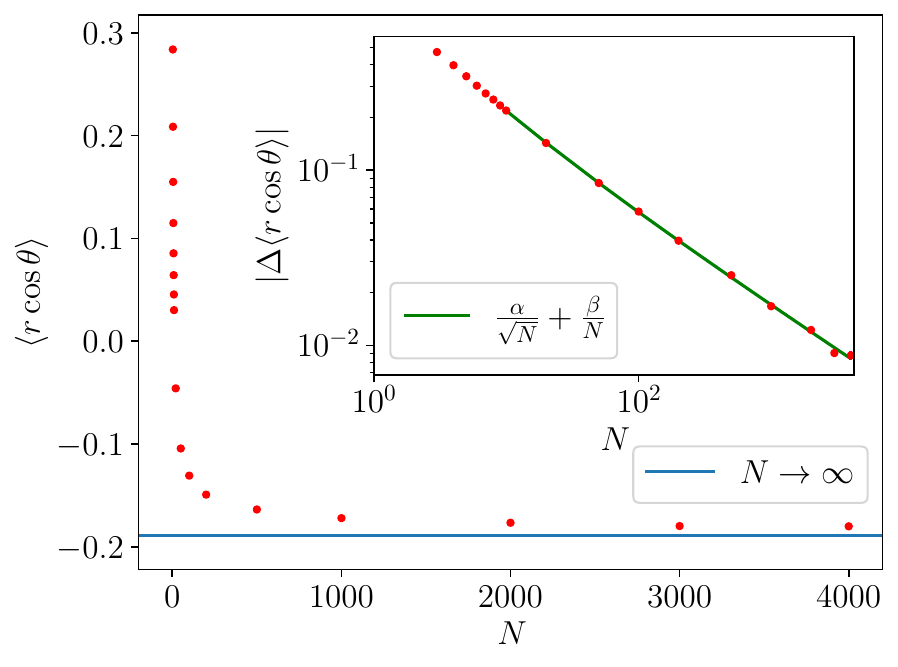} \\[-1mm] 
  \end{tabular}
  \caption{Similar to \cref{fig:moments_analytical}, but with moments obtained from the numerical diagonalization of $N$-dimensional random matrices from the Ginibre unitary ensemble. For every $N$ the number of random matrices was chosen to obtain a total of $10^6$ complex spacings. If the error bars are not visible they are smaller than the symbols. The curves in the insets are fits to the function $\alpha/\sqrt N+\beta/N$ in the range $N\in[10,4000]$.}
  \label{fig:moments_rmt}
\end{figure*}

We now investigate how the moments computed for $N$-dimensional random matrices from the Ginibre unitary ensemble behave as a function of $N$. The results are shown in \cref{fig:moments_rmt}. We observe that the convergence to the $N\to\infty$ limit is much slower than for our approximation formula and asymptotically goes like $1/\sqrt N$. As already observed in \cite{PhysRevX.10.021019} and confirmed in \cite{garciagarcia2021symmetry}, this is largely due to boundary effects. Indeed, if the $N$ eigenvalues occupy a certain area in the complex plane, the ratio of boundary to area generically scales like $1/\sqrt N$.

The convergence of the marginal distributions (not shown) to the $N\to\infty$ limit is also slow. Thus, when studying a particular physical system with complex eigenvalues, it is advisable to compare the results for the spacing ratios to our approximation formula rather than to data obtained from diagonalizing random matrices.

\section{Summary}
\label{sec:summary}

As pointed out in \cite{PhysRevX.10.021019}, complex spacing ratios are an important tool in the analysis of non-Hermitian systems because they eliminate the need for unfolding and can unambiguously discriminate between regular and chaotic dynamics. We have derived an approximation formula for the distribution $P(z)$ of complex spacing ratios in the Ginibre universality class that converges exponentially fast to the $N\to\infty$ limit. In contrast, the exact finite-$N$ results converge to this limit only like $1/\sqrt N$.
Therefore, our result appears to be more suitable for comparison with empirical data from large systems.

Our approximations to $P(z)$ and the corresponding marginal distributions $P(r)$ and $P(\theta)$ are given in terms of simple power series. The coefficients of the power series and the moments of the distributions can be obtained from our computer algebra code \cite{csratio}.
In future work it would be interesting to perform similar calculations for the other two universality classes, \AId and \AIId. This is a more complicated problem since, to the best of our knowledge, the joint probability distribution of the eigenvalues is not known yet for these ensembles. 

\bibliography{references}

\end{document}